\documentclass{webofc}
\usepackage[varg]{txfonts}   % Web of Conferences font
\usepackage{hyperref}
\usepackage{url}
\usepackage{lineno}
\hypersetup{colorlinks=true,citecolor=black,urlcolor=black,linkcolor=black}

\begin{document}
%
%\linenumbers

\title{Evidence for modest octupole deformation in $^{238}$U from high-energy heavy-ion collisions}

\author{\firstname{Chunjian} \lastname{Zhang}\inst{1,2}\fnsep\thanks{\email{chunjianzhang@fudan.edu.cn}} \lastname{(for the STAR Collaboration)} 
}

\institute{
Key Laboratory of Nuclear Physics and Ion-beam Application (MOE), and Institute of Modern Physics, Fudan
University, Shanghai 200433, China
\and
Shanghai Research Center for Theoretical Nuclear Physics, NSFC and Fudan University, Shanghai 200438, China
}

\abstract{We present a novel ``imaging-by-smashing" approach for probing nuclear deformation in high-energy heavy-ion collisions. By analyzing anisotropic-flow ($v_n$) and mean transverse momentum ($\left[p_T\right]$)-based observables in collisions of highly deformed $^{238} \mathrm{U}$ nucleus and nearly spherical $^{197} \mathrm{Au}$ nucleus, we extract the deformation parameters of $^{238}$U. The key observables include the variances $\left\langle v_n^2\right\rangle,\left\langle\left(\delta p_T\right)^2\right\rangle$, and the covariance $\left\langle v_n^2 \delta p_T\right\rangle$. Ratios of these observables between $^{238}$U+$^{238}$U and $^{197}$Au+$^{197}$Au collisions largely cancel final-state effects, thereby isolating the influence of nuclear deformation. We further report the first experimental indication of octupole deformation in $^{238}$U via $v_3$-based observables~\cite{2025rot}. The extracted deformation parameters, comparing with state-of-the-art hydrodynamic model calculations, are consistent with low-energy nuclear structure data. These results establish high-energy collisions as a powerful probe of nuclear shapes on femtosecond timescales.
}
\maketitle
\section{Introduction}
\label{intro}
Understanding the initial conditions and transport properties of the deconfined quark–gluon plasma (QGP) produced in high-energy heavy-ion collisions is a major focus of high energy nuclear research. One powerful way to constrain these initial conditions is to determine the distinct nuclear structures of the colliding nuclei, thereby providing valuable insights into the initial geometry and the subsequent energy deposition mechanisms~\cite{Jia:2022ozr}. Traditionally, nuclear shapes are inferred from spectroscopic or scattering experiments conducted at sub-hundred MeV/nucleon beam energies, probing collective degrees of freedom on timescales longer than 100 fm/$c$. In contrast, high-energy nuclear collisions are strongly interacting, highly dynamical systems, for which the stationary solution of an individual nucleus is not an eigenstate of the combined two-nuclei system. Such collisions are sensitive not only to the global nuclear shapes relevant at low-energy, but also to fluctuations of the nuclear wavefunction on much shorter timescales~\cite{Jia:2022ozr,Duguet:2025hwi,Ke:2025tyv}. This makes high-energy collisions a novel tool for investigating nuclear structure and its evolution with energy. 

\section{``Imaging-by-smashing" method}\label{sec1}
We introduce a novel ``imaging-by-smashing" technique: a collective-flow-assisted nuclear shape-imaging method, that extracts an instantaneous ``snapshot" of the nuclear geometry from the collective response of the debris in ultra-relativistic heavy-ion collisions~\cite{STAR:2024wgy,Giacalone:2021udy,Zhang:2021kxj,Jia:2022qgl,Jia:2025wey}. Nuclear deformation affects such collisions by modifying the nucleon distribution in the overlap region and thereby altering the initial conditions of the QGP. Random orientations of deformed nuclei lead to substantial event-by-event variations in both the initial size and shape of the QGP fireball, which in turn enhance fluctuations in the resulting anisotropic and radial flow.

Nuclear shape is described by a Woods-Saxon density profile,
\begin{equation}\begin{split}
\begin{aligned}
\label{eq:1}
\rho(r, \theta, \phi) \propto(1+\exp [r-R(\theta, \phi) / a])^{-1}, \\
R(\theta, \phi)=R_0\left(1+\beta_2\left[\cos \gamma Y_{2,0}+\sin \gamma Y_{2,2}\right]+\beta_3 Y_{3,0}+\beta_4 Y_{4,0}\right),
\end{aligned}
\end{split}
\end{equation}
where $R_0=1.2 A^{1/3}$ is the nuclear radius, $A$ is the nuclear mass number, $a$ is the surface or skin depth, and $Y_{l, m}(\theta, \phi)$ are real spherical harmonics in the intrinsic frame. The axial-symmetric quadrupole, octupole and hexadecapole deformations are denoted by $\beta_2$, $\beta_3$, and $\beta_4$, respectively. The triaxiality parameter $\gamma$ ($0^{\circ} \leq \gamma \leq 60^{\circ}$) determines the relative ordering of the three principal radii.

For small deformations, a Taylor expansion shows that $v_n$-based observables follow simple parametric dependencies on the nuclear shape parameters: 
$\left\langle v_2^2\right\rangle =a_1+b_1 \beta_2^2$, 
$\left\langle v_3^2\right\rangle =a_3+b_3 \beta_3^2$, 
$\left\langle\left(\delta p_{\mathrm{T}}\right)^2\right\rangle=a_2+b_2 \beta_2^2$, $\left\langle v_2^2 \delta p_{\mathrm{T}}\right\rangle =a_3-b_3 \beta_2^3 \cos (3 \gamma)$, 
$\left\langle v_3^2 \delta p_{\mathrm{T}}\right\rangle =a_4-b_4 \beta_2 \beta_3^2$ (See Refs.~\cite{Jia:2022ozr,Zhang:2021kxj,STAR:2024wgy}). The positive coefficients $a_n$ and $b_n$, are determined by the collision geometry and the properties of the QGP. While the $b_n$ values are nearly centrality independent, the $a_n$ values  reach their minimum in central collisions, making such collisions events particularly sensitive for constraining nuclear shape (see Refs.~\cite{Giacalone:2021udy,Jia:2022ozr}). 

\section{Results}\label{sec2}
\begin{figure}[ht]
\sidecaption
\centering
\includegraphics[width=0.49\linewidth]{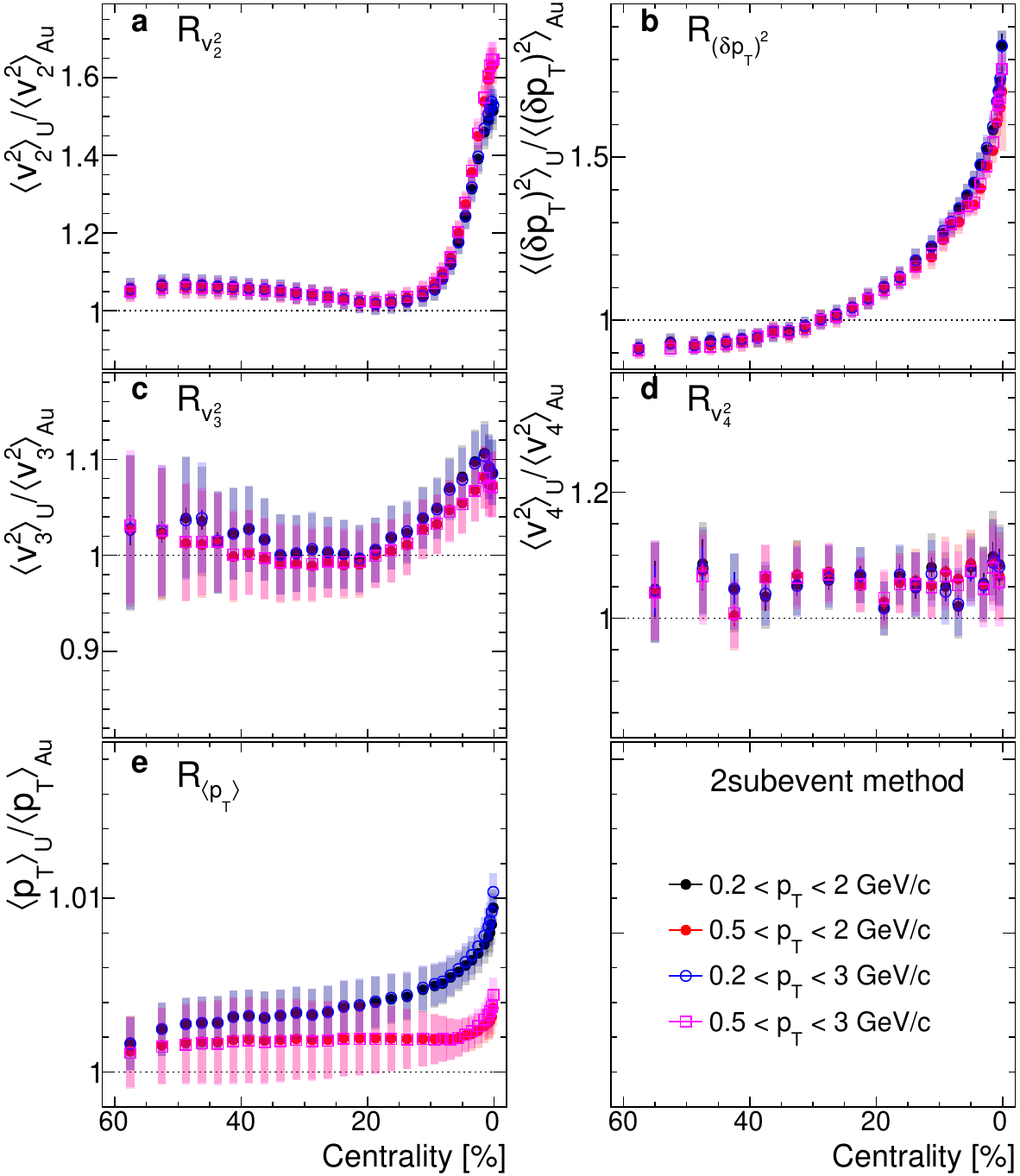}
\caption{Centrality dependence of ratios of one-and two-particle observables, $\left\langle v_2^2\right\rangle,\left\langle\left(\delta p_{\mathrm{T}}\right)^2\right\rangle$, $\left\langle v_3^2\right\rangle,\left\langle v_4^2\right\rangle$, and $\left\langle p_{\mathrm{T}}\right\rangle$, between $^{238}\mathrm{U}+^{238}\mathrm{U}$ and $^{197}\mathrm{Au}+^{197}\mathrm{Au}$ collisions in four $p_{\mathrm{T}}$ ranges. Figure is from Ref.~\cite{2025rot}.}
\label{fig1}       % Give a unique label
\end{figure}

We measure the correlators $\langle p_T \rangle$, $\left\langle\left(\delta p_{\mathrm{T}}\right)^2\right\rangle$, $\left\langle v_n^2\right\rangle$, and $\left\langle v_n^2 \delta p_{\mathrm{T}}\right\rangle$.
Direct ratios of the same observable between $^{238}\mathrm{U}+^{238}\mathrm{U}$ and $^{197}\mathrm{Au}+^{197}\mathrm{Au}$ collisions, defined as $R_{\mathcal{O}}=\frac{\langle\mathcal{O}\rangle_{\mathrm{U}}}{\langle\mathcal{O}\rangle_{\mathrm{Au}}}$ at matched centrality, suppress influences from initial-state properties common to both systems, such as nucleon sizes, nucleon distances, and nucleon substructure fluctuations. Figure~\ref{fig1} shows the resulting ratios of soft probes. Although the absolute magnitudes of $\left\langle v_n^2\right\rangle$ and $\left\langle\left(\delta p_{\mathrm{T}}\right)^2\right\rangle$ vary by more than a factor of two across the four $p_{\mathrm{T}}$ intervals, their ratios exhibit a striking $p_{\mathrm{T}}$ independence, indicating significant cancellation of final-state effects. In central collisions, $R_{v_2^2}$ and $R_{\left(\delta p_{\mathrm{T}}\right)^2}$ exhibit a pronounced enhancement, reflecting the larger quadrupole deformation of $^{238}$U. The enhancement appears over a narrow centrality interval for $R_{v_2^2}(0 \%-10 \%)$ but spans a much broader range for $R_{\left(\delta p_{\mathrm{T}}\right)^2}(0 \%-30 \%)$. This difference arises because $\left\langle v_2^2\right\rangle$ in non-central collisions is dominated by the elliptic geometry of the overlap region, thereby diluting the impact of nuclear deformation, whereas $\left\langle\left(\delta p_{\mathrm{T}}\right)^2\right\rangle$ is unaffected by this average geometric contribution. 
We observe $\left\langle v_3^2\right\rangle$ values in central collisions are about $7 \%$ larger in $\mathrm{U}+\mathrm{U}$ compared to $\mathrm{Au}+\mathrm{Au}$, as shown in $R_{v_3^2}$. Naively, we expect the $\mathrm{Au}+\mathrm{Au}$ system, being smaller, to have larger fluctuations in its initial geometry (specifically, larger $\left\langle\varepsilon_3^2\right\rangle$ ). This simple scaling, $\left\langle\varepsilon_3^2\right\rangle \propto 1 / A$, suggests that $\left\langle v_3^2\right\rangle_{\mathrm{Au}}$ should be greater than $\left\langle v_3^2\right\rangle_{\amalg}$. The surprising enhancement of $R_{v_3^2}$ values is possibly from octupole deformation in $^{238}$U, while $R_{v_4^2}$ shows no centrality-dependent modification.

With sufficient statistics, the $\left\langle p_{\mathrm{T}}\right\rangle$ ratio exhibits a modest ($\sim 1 \%$) increase toward central collisions, possibly reflecting the denser medium produced in $^{238}\mathrm{U}+^{238}\mathrm{U}$ collisions and the bias toward tip-tip configurations in UCC (ultra-central collision) events~\cite{STAR:2015mki}. The choice of $p_{\mathrm{T}}$ interval affects the absolute value of $\left\langle p_{\mathrm{T}}\right\rangle$ with increasing lower $p_{\mathrm{T}}$ limits naturally lowering $\left\langle p_{\mathrm{T}}\right\rangle$. This kinematic effect weakens the enhancement observed in the UCC region, leading to a separation of the ratios into two groups determined by the lower $p_{\mathrm{T}}$ threshold.  In contrast, varying the upper $p_T$ limit produces little change because particle yields at high $p_T$ are comparatively small.

\begin{figure}[ht]
\sidecaption
\centering
\vspace*{1cm} 
\includegraphics[width=0.5\linewidth]{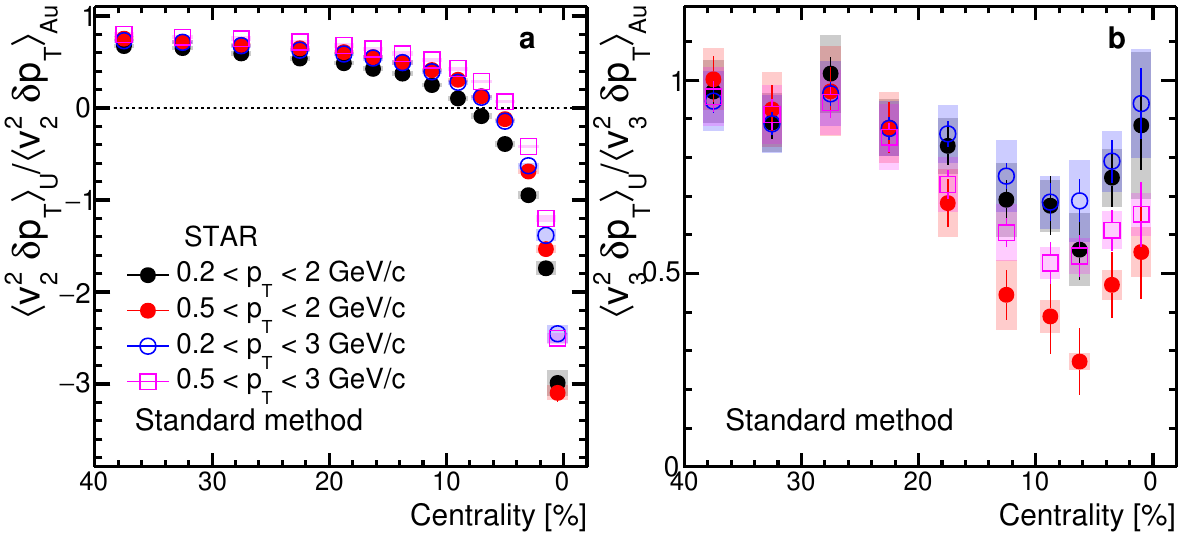}
\caption{Centrality dependence of ratios of three-particle observable, $\left\langle v_2^2 \delta p_{\mathrm{T}}\right\rangle$ (left) and $\left\langle v_3^2 \delta p_{\mathrm{T}}\right\rangle$ (right) between $^{238}\mathrm{U}+^{238}\mathrm{U}$ and $^{197}\mathrm{Au}+^{197}\mathrm{Au}$ collisions, obtained for four $p_{\mathrm{T}}$ intervals. Figure is from Ref.~\cite{2025rot}.}
\label{fig2}       % Give a unique label
\end{figure}

Figure~\ref{fig2} presents the ratios of three-particle correlator $\left\langle v_n^2 \delta p_{\mathrm{T}}\right\rangle$ between $\mathrm{U}+\mathrm{U}$ and $\mathrm{Au}+\mathrm{Au}$ collisions for four $p_{\mathrm{T}}$ intervals. The ratios $R_{v_2^2 \delta p_{\mathrm{T}}}$ are essentially independent of the chosen $p_{\mathrm{T}}$ interval across all centralities, whereas $R_{v_3^2 \delta p_{\mathrm{T}}}$ exhibits a modest dependence on $p_T$ selection in the UCC region.

Many of the measured ratios exhibit significant deviations from
unity in the UCC region. These deviations can be used to constrain the quadrupole and higher-order deformation of $^{238}$U when compared with detailed hydrodynamic and initial-state model calculations. We compare the experimental ratios with state-of-the-art IP-Glasma+MUSIC and Glauber model results. By varying-medium evolution and initial-state parameters, and incorporating nonflow effects in the data, we extract $\beta_{2 \mathrm{U}}=0.286 \pm 0.025$ and $\gamma_{\mathrm{U}}=8.7 \pm 4.5^{\circ}$~\cite{STAR:2024wgy}. These values are broadly consistent with low-energy estimates under the rigid-rotor assumption~\cite{Pritychenko:2013gwa}. In addition, we also observe evidence for high-order deformations in the ground states of $^{238}$U for the first time. The IP-Glasma+MUSIC calculations support a modest octupole deformation, $\beta_{3\rm U} \sim 0.08-0.10$~\cite{Zhang:2025hvi}. Directly measuring higher-order deformations of low-lying nuclear ground states is particularly crucial yet experimentally challenging. This constitutes the first experimental indication, at high energies, of octupole deformation in $^{238}$U. Notably, these constraints are derived without reference to transitions involving excited nuclear states, illustrating that ultra-central high-energy collisions provide direct sensitivity to ground-state nuclear shapes.

\section{Summary and outlook}
In summary, we presented the ratios of several observables, $\langle p_T \rangle$, $\left\langle\left(\delta p_{\mathrm{T}}\right)^2\right\rangle$, $\left\langle v_n^2\right\rangle$, and $\left\langle v_n^2 \delta p_{\mathrm{T}}\right\rangle$, between $^{238}\mathrm{U}+^{238}\mathrm{U}$ and $^{197}\mathrm{Au}+^{197}\mathrm{Au}$ collisions in four $p_T$ intervals. Significant deviations of these ratios from unity are observed in these observables, indicating sensitivity to features of nuclear structure. By comparing the data with hydrodynamic and initial-state model calculations, we determine $\beta_{2 \mathrm{U}}=0.286 \pm 0.025$ and $\gamma_{\mathrm{U}}=8.7 \pm 4.5^{\circ}$. We further report the first experimental indication from the high-energy collisions for octupole collectivity in $^{238}$U, corresponding to a modest $\beta_{\rm 3U} \sim 0.08-$ 0.10. These values are broadly consistent with low-energy estimates obtained under the rigid-rotor assumption, demonstrating that high-energy collisions can provide a valuable quantitative determination of the nuclear ground-state shape.  

Imaging nuclear shapes through high-energy heavy-ion collisions offers a promising avenue for constraining  the elusive QGP initial conditions and for probing the evolution of exotic nuclear structure with energy~\cite{Ma:2022dbh,Giacalone:2025vxa}. Further studies, including the intermediate isobaric $^{96}$Ru+$^{96}$Ru and $^{96}$Zr+$^{96}$Zr collisions, and light-ion $^{16}$O+$^{16}$O collisions at RHIC, will futher help to refine this ``imaging-by-smashing" technique. With continued proper validation, this method can become a powerful and complementary tool for investigating collective many-body nuclear structure across the nuclide chart and over a broad range of energy scales. 
\\
\noindent $\boldsymbol{\rm Acknowledgment:}$ This work is supported in part by the National Key Research and Development Program of China under Contract Nos. 2024YFA1612600 and 2022YFA1604900, the National Natural Science Foundation of China (NSFC) under Contract Nos. 12025501, 12147101, 12205051, the Natural Science Foundation of Shanghai under Contract No. 23JC1400200, Shanghai Pujiang Talents Program under Contract No. 24PJA009, the U.S. Department of Energy, Office of Science, Office of Nuclear Physics, under Award No. DE-SC0024602.
\bibliography{reference}{}

\end{document}